\documentclass[aps,prb,twocolumn]{revtex4}
\usepackage{graphicx}
\usepackage{amssymb,amsmath}
\usepackage{subfigure}

\begin{document}
\bibliographystyle{apsrev}

\title{Structure and Diffusion of Nanoparticle Monolayers
Floating at Liquid/Vapor Interfaces: A Molecular Dynamics Study}
\author{Shengfeng Cheng}
\email{sncheng@sandia.gov}
\affiliation{Sandia National Laboratories, Albuquerque, NM 87185, USA}
\author{Gary S. Grest}
\affiliation{Sandia National Laboratories, Albuquerque, NM 87185, USA}

\date{\today}

\begin{abstract}
Large-scale molecular dynamics simulations are used to simulate 
a layer of nanoparticles diffusing on the surface of a liquid. 
Both a low viscosity liquid, represented by Lennard-Jones monomers, 
and a high viscosity liquid, represented by linear homopolymers, are studied. 
The organization and diffusion of the nanoparticles are analyzed 
as the nanoparticle density and the contact angle between the nanoparticles and liquid are varied. 
When the interaction between the nanoparticles and liquid is reduced 
the contact angle increases and the nanoparticles ride higher on the liquid surface, 
which enables them to diffuse faster. 
In this case the short range order is also reduced as seen in the pair correlation function. 
For the polymeric liquids, the out-of-layer fluctuation 
is suppressed and the short range order is slightly enhanced.
However, the diffusion becomes much slower and the mean square displacement even
shows sub-linear time dependence at large times.
The relation between diffusion coefficient and viscosity is found to deviate
from that in bulk diffusion.
Results are compared to simulations of the identical nanoparticles in 2-dimensions.
\end{abstract}


\maketitle

\noindent{\bf I. INTRODUCTION}
\bigskip

Nanoparticles at a liquid/vapor or liquid/liquid interface have attracted
extensive attention during the past two decades.\cite{bresme07}
One motivation is that nanoparticles adsorbed at interfaces are found
to be able to stabilize emulsions and foams.\cite{binks00,binks02,horozov06}
Nanoparticles also self-assemble into various structures 
at an interface, which provides an efficient route to produce
superlattices of nanoparticles of technological importance.\cite{lin03,binder05,boker07} 
The advantage of this technique is that the assembly process can be fast 
and the resulting arrays are usually highly ordered.
Due to their small size 
the adsorption energy of nanoparticles at an interface
is typically only a few $k_{\rm B}T$, where $k_{\rm B}$ is the Boltzmann constant and $T$
is the temperature.\cite{lin03,bresme07,du10}
The small adsorption energy implies that the adsorbed nanoparticles 
are highly dynamic and the self-assembly is quite reversible, 
the latter of which has been found to be crucial to form highly ordered arrays.\cite{whitesides02,binder05}.
Nanoparticles straddling an interface can also be regarded as 
living in a quasi 2-dimensional (2D) environment, which provides a model system to study 
interesting problems such as phase transitions 
of 2D fluids.\cite{pieranski80,terao99,zahn00,sun03}

Many experimental studies have been devoted to 
investigate the factors controlling the behavior of nanoparticles
at an interface, including their size, surface morphology, shape, materials polarizability, 
and coatings.\cite{heath97,glaser06,isa11,zang11,comeau12} 
Some of these factors influence the location and orientation of individual nanoparticles; 
others influence their mutual interactions and assembly geometry. 
However, in experiments these factors are usually intertwined to yield collective effects
and it is difficult to single out the effect of each factor alone. 
This aspect is where molecular dynamics (MD) simulations can play a useful role by 
studying the effect of one factor at a time to help 
elucidate experimental observations and uncover new physical 
insights.\cite{bresme98,fenwick01,powell02,bresme09,chiu09,cheung10,cheung11,fan11,frost11}
For example, Bresme {\it et al.} showed that Young's equation can 
be used to describe force balance at nanoscale interfaces,
but in certain cases line tension should also be included.\cite{bresme98}
Fenwick and Powell {\it et al.} showed that contrary to expectations, 
the collapse pressure measured in a typical Langmuir trough experiment should be
independent of the contact angle.\cite{fenwick01,powell02}
Recently, Cheung showed the importance of nanoparticle-liquid interactions and 
capillary waves in determining the
stability of nanoparticles at liquid interfaces.\cite{cheung11}

\begin{figure}[ht]
  \subfigure[]{
    \begin{minipage}[b]{0.15\textwidth}
      \centering
      \includegraphics[width=1in]{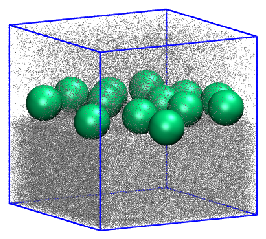}
     \end{minipage}}%
  \subfigure[]{
    \begin{minipage}[b]{0.15\textwidth}
      \centering
      \includegraphics[width=1in]{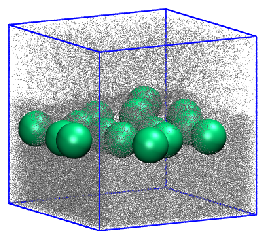}
     \end{minipage}}%
  \subfigure[]{
    \begin{minipage}[b]{0.15\textwidth}
      \centering
      \includegraphics[width=1in]{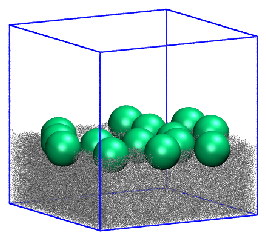}
     \end{minipage}}
\caption{(Color online) 
Snapshots of nanoparticles floating at liquid/vapor interfaces:
(a) monatomic LJ liquid and $\theta_c=137^\circ$;
(a) monatomic LJ liquid and $\theta_c=29^\circ$;
(c) 100-bead chain polymeric liquid and $\theta_c=93^\circ$.
Only a small portion ($100\sigma \times 100\sigma \times 100\sigma$) 
of the simulation cell is shown in each snapshot.
}
\label{FigSnapshot}
\end{figure}

Although MD simulations to date have revealed many important aspects of the physical behavior
of nanoparticles at an interface, some aspects are still unclear.
Particularly, it is not clear how the structure and dynamics depend on contact angle and nanoparticle
density in the low coverage regime. 
These behaviors are important since low density clusters can occur at the earlier stage
of assembly and affect the morphology of the final dense layer.\cite{bigioni06} 
The effect of liquid viscosity on interfacial diffusion is also not well understood 
and has only been studied recently.\cite{cheung10,wang11}
In this paper we use MD to study the structure and dynamics of a layer of 
nanoparticles floating at a liquid/vapor interface. 
We focus on the effect of varying the contact angle,
which is controlled by interactions between the nanoparticles and liquid, 
the nanoparticle density, and the liquid viscosity, respectively.
Snapshots of some of the systems we have simulated are shown in 
Fig.~\ref{FigSnapshot} (only a part of the full simulation box is shown in each case).
In Fig.~\ref{FigSnapshot}(a)  the contact angle between the nanoparticle and liquid 
is $\theta_c=137^\circ$. In this case the nanoparticles
only slightly dip into the liquid which is composed of Lennard-Jones (LJ) monomers
and is in equilibrium with its vapor phase. In
Fig.~\ref{FigSnapshot}(b) $\theta_c=29^\circ$ and the 
nanoparticles are almost immersed in the liquid.
In Fig.~\ref{FigSnapshot}(c) $\theta_c=93^\circ$ and the 
nanoparticles straddle the surface of the liquid, in this case composed
of flexible linear polymer chains. 
Note that the LJ monomer liquid has a high vapor density,
but the vapor density of the polymeric liquid is essentially $0$.\cite{cheng11}
Also note that the polymeric liquid has a higher bulk density than that of the monatomic liquid 
and the liquid film in Fig.~\ref{FigSnapshot}(c) is thinner 
than those in Fig.~\ref{FigSnapshot}(a) and (b) since in our simulations 
all liquids contain a similar number of monomers.

\bigskip\noindent{\bf II. SIMULATION METHODOLOGY}
\bigskip

We placed a layer of nanoparticles at a liquid/vapor interface 
as shown in Fig.~\ref{FigSnapshot}. 
Three liquid systems consisting of either LJ monomers 
or flexible linear chains of $N$ LJ beads for $N=10$ and $100$ were studied.
In all three cases, the beads interact
with each other through the standard LJ 12-6 potential
\begin{equation}
U_{\rm LJ}(r)=4\epsilon\left[ \left(\frac{\sigma}{r}\right)^{12}-\left(\frac{\sigma}{r}\right)^6
-\left(\frac{\sigma}{r_c}\right)^{12}+\left(\frac{\sigma}{r_c}\right)^6 \right],
\label{LJPotential}
\end{equation}
where $r$ is the distance between two beads, 
$\epsilon$ is the unit of energy, and $\sigma$ is the diameter of beads. 
The interaction is truncated at $r_c=3.0\sigma$.
For the two polymeric liquids, beads in the
chain are connected by an additional finite extensible
nonlinear elastic potential with a spring
constant $k=30\epsilon/\sigma^2$ and 
maximum extent $R_0 =1.5\sigma$.\cite{kremer90}
The liquid density $\rho$ and shear viscosity $\eta$ for the LJ monomer liquid
is $\rho= 0.66\sigma^{-3}$ and $\eta=1.01\pm 0.03 m/\tau\sigma$,\cite{petersen10b} while
for the polymer liquids, $\rho=0.89\sigma^{-3}$ and $\eta=12\pm 1 m/\tau\sigma$
for $N=10$, and $\rho=0.91\sigma^{-3}$ and $\eta=180\pm 10 m/\tau\sigma$ 
for $N=100$.\cite{note1,tuckerman97}

The nanoparticles are assumed to consist of a uniform
distribution of atoms interacting with a LJ potential.
For spherical particles, their mutual interaction
can then be determined analytically
by integrating over all the interacting LJ atom pairs
between the two particles.\cite{Hamaker37,Everaers03}
For nanoparticles with radii $a$, the interaction energy is given by 
\begin{equation}
\label{ColPotential}
\begin{array}{lll}
U_{\rm nn}(r)& = -\frac{A_{\rm nn}}{6}\left[ 
      \frac{2 a^2}{r^2 - 4a^2} +
      \frac{2 a^2}{r^2} +
      {\rm ln}\left( \frac{r^2 - 4a^2}{r^2} \right)\right]\\
 & +\frac{A_{\rm nn}}{37800}\frac{\sigma_n^6}{r}\left[
    \frac{r^2-14ra+54a^2}{(r-2a)^7}\right.\\
 & \left. +\frac{r^2+14ra+54a^2}{(r+2a)^7}
	-\frac{2(r^2-30a^2)}{r^7}\right].
\end{array}
\end{equation}
Here $r$ is the center-to-center distance between two nanoparticles.
The Hamaker constant $A_{\rm nn}=4\pi^2\epsilon_{\rm nn}\rho_n^2 \sigma_n^6$, where 
$\epsilon_{\rm nn}$ is the interaction strength between the LJ atoms 
that make up the nanoparticles, and $\sigma_n$ is the diameter 
and $\rho_n$ the density of LJ atoms in the nanoparticles. 
To reduce the number of parameters, we take $\epsilon_{\rm nn}=\epsilon$, 
$\sigma_n=\sigma$ and $\rho_n=1.0\sigma^{-3}$, 
in which case $A_{\rm nn}=39.48\epsilon$. 
In this paper we set $a=10\sigma$.

The interaction between the LJ beads and
nanoparticles is determined by integrating the interaction
between a LJ bead and the LJ atoms within a nanoparticle,
and the interaction potential $U_{\rm ns}(r)$ is given by 
\begin{equation}
\begin{array}{lll}
U_{\rm ns}(r)& = \frac{2}{9}\frac{a^3\sigma_{n}^3 A_{\rm ns}}{(a^2-r^2)^3}\\
 & \times \left[1-\frac{(5a^6+45a^4r^2+63a^2r^4+15r^6)\sigma_{n}^6}
{15(a^2-r^2)^6}\right],
\end{array}
\label{ColSolPotential}
\end{equation}
where $r$ is the center-to-center distance between the bead and nanoparticle,
and the Hamaker constant $A_{\rm ns}=24\pi\epsilon_{\rm ns}\rho_n \sigma_n^3=24\pi\epsilon_{\rm ns}$ 
for $\sigma_{n}=\sigma$ and $\rho_n=1.0\sigma^{-3}$.

\begin{figure}[htb]
\centering
\includegraphics[width=2.75in]{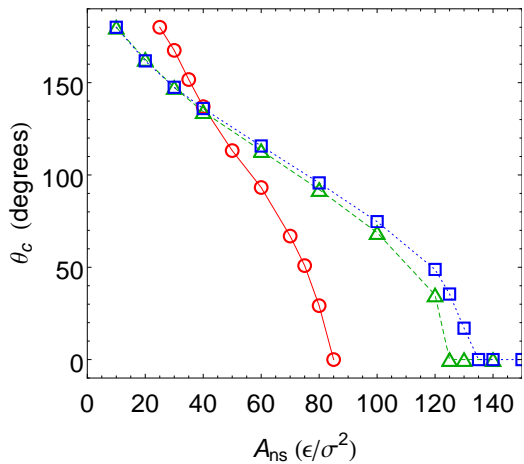}
\caption{(Color online)
Contact angle $\theta_c$ vs. nanoparticle-liquid 
interaction strength $A_{\rm ns}$ for different liquids:
monatomic LJ liquid (circles),
10-bead chain polymeric liquid (triangles),
and 100-bead chain polymeric liquid (squares).
Uncertainties in $\theta_c$ are $\sim 1^\circ-2^\circ$,
smaller than the symbol size. Lines are guides to the eye.}
\label{FigConAng}
\end{figure}

Depending on the values of the Hamaker constant $A_{\rm nn}$ and $A_{\rm ns}$, the
nanoparticles can either be dispersed in the liquid or phase separate. 
As we are interested in studying nanoparticles that are hard-sphere like 
at the liquid/vapor surface, we truncate the 
nanoparticle-nanoparticle interaction so that it is purely repulsive. 
For $a=10\sigma$, this gives a cutoff $r_c=20.427\sigma$ for $U_{\rm nn}(r)$. 
Physically this corresponds to adding a short surfactant
coating to the nanoparticles to avoid flocculation.\cite{veld09,grest11} 
For the interaction between the nanoparticles and LJ beads making up the liquid and vapor, 
we set $r_c=a+4\sigma=14\sigma$. The remaining free parameter 
$A_{\rm ns}$ controls
the solubility of the nanoparticles in the liquid.
We choose $A_{\rm ns}$ such that nanoparticles phase separate 
to the liquid/vapor interface, in which case
$A_{\rm ns}$ controls the contact angle $\theta_c$ of 
the nanoparticles on the liquid surface.
Results for $\theta_c$ as a function of $A_{\rm ns}$ are shown in Fig.~\ref{FigConAng}.
The data show that $\theta_c \rightarrow 180^\circ$ as $A_{\rm ns} \rightarrow 0$.
However, when $A_{\rm ns}$ exceeds certain critical value, which
is approximately $85\epsilon$ for LJ monomers and $120\epsilon$ ($130\epsilon$) 
for 10-bead (100-bead) chains, 
$\theta_c$ goes to $0$ and the nanoparticles diffuse into
the liquid. In the intermediate range around $90^\circ$, 
$\theta_c$ decreases roughly
linearly as $A_{\rm ns}$ increases.
For a fixed $A_{\rm ns}$, $\theta_c$ increases
as the chain length increases. This trend can be qualitatively understood as
the result of less entropy gain when mixing the nanoparticles with
longer chains.
Correspondingly, a larger $A_{\rm ns}$, which represents
the enthalpy contribution of solvation,
is required to disperse the nanoparticles into a liquid
of longer chains.\cite{note3}
The effects of $A_{\rm ns}$ and chain length 
on the behavior of the nanoparticles at the liquid/vapor interface 
are visualized in Fig.~\ref{FigSnapshot}.

All MD simulations were performed using the LAMMPS simulation package.\cite{plimpton95,lammps}
The simulation cell is a rectangular box of dimensions $L_x \times L_y \times L_z$, 
where $L_x=390.577\sigma$, $L_y=451.0\sigma$, and $L_z=120\sigma$
for the LJ monomer liquid and $70\sigma$ for the polymeric liquids.
The liquid/vapor interface is parallel to the $x$-$y$ plane,
in which periodic boundary conditions were employed.
In the $z$ direction, the LJ atoms and nanoparticles are confined 
by two flat walls at $z=0$ and $z=L_z$, respectively.
Since the 2D packing of hard spheres is very sensitive to the aspect ratio
of the enclosing box,
we set $L_x/L_y=\sqrt{3}/2$ so that the hexagonal close-packing is favored.
Each system contains more than $6$ million LJ atoms to form a liquid layer with a thickness 
$\sim 50\sigma$ and in equilibrium with its vapor phase, which in the LJ monomer case
has a thickness $\sim 70\sigma$ for $L_z=120\sigma$. 
For the polymeric liquids, the vapor density is $0$ and $L_z$ was reduced accordingly.
In order to investigate the effect of nanoparticle coverage,
three systems were simulated with 
$N_p = 200$, $240$, and $320$ nanoparticles, 
corresponding to 2D density $\phi\equiv N_p(2a)^2/(L_xL_y)=0.45$, $0.54$, and $0.73$, respectively.
All these densities are well below the critical density $0.89$ 
at the liquid/hexatic transition and $0.92$ at 
the hexatic/solid transition of 2D hard disk fluids.\cite{alder62,binder02,mak06}

The LJ atoms interact with both upper and lower walls through 
a LJ 9-3 potential, which depends only on their distance $z$ from the wall,
\begin{equation}
U(z)=\epsilon_{\rm w}\left[ \frac{2}{15}\left(\frac{\sigma}{z}\right)^{9}
-\left(\frac{\sigma}{z}\right)^3 
-\frac{2}{15}\left(\frac{\sigma}{z_c}\right)^{9}
+\left(\frac{\sigma}{z_c}\right)^3 \right],
\label{WallPotential}
\end{equation}
where $\epsilon_{\rm w} = 2\epsilon$. 
For the lower wall the interaction is truncated at $z_c=3.0\sigma$, 
while at the upper wall is purely repulsive with $z_c=0.71476\sigma$. 
Though all nanoparticles are confined to the liquid/vapor interface and 
far from the two walls in most simulations, we also included a 
nanoparticle-wall potential of the form
\begin{equation}
\begin{array}{lll}
U(z)& = A_{\rm nw}\left[ \frac{\sigma^6}{7560}\left( 
\frac{7a-z}{(z-a)^7}+\frac{7a+z}{(z+a)^7}\right)\right.\\
 & \left.-\frac{1}{6}\left( \frac{2az}{z^2-a^2} 
+ {\rm ln}\frac{z-a}{z+a} \right) \right],
\end{array}
\label{ColWallPotential}
\end{equation}
where $z$ is the distance of the center of a nanoparticle from the wall and
$A_{\rm nw}=144\epsilon$. At both walls the potential
is truncated at $z=10.57187\sigma$ to make it purely repulsive for the nanoparticles.
This potential is useful for larger values of $A_{\rm ns}$ 
where the nanoparticles are dispersed in the liquid.\cite{grest11}

The equations of motion were integrated using a velocity-Verlet algorithm 
with a time step $\delta t =0.005\tau$,
where $\tau=\sigma(m/\epsilon)^{1/2}$ and $m$ is the mass of a LJ monomer.
The nanoparticle with a radius $a=10\sigma$
has a mass $M=\frac{4 \pi a^3 m}{3\sigma^3}=4188.79m$.
During the equilibration, the temperature 
$T$ was held at $1.0\epsilon/k_{\rm B}$ by a Langevin thermostat weakly 
coupled to all LJ atoms with a damping constant $\Gamma=0.1\tau^{-1}$. 
Once the liquid/vapor interface was equilibrated, 
the Langevin thermostat was removed except
for those liquid atoms within $10\sigma$ 
of the lower wall at $z=0$.
Since the thickness of the liquid layer is $\sim 50\sigma$ and
all nanoparticles are floating at the liquid/vapor interface in our simulations, 
their motion is not affected by the thermostat.

For comparison, we also
conducted MD simulations of 5000 nanoparticles in a 2D box
with $L_x/L_y=1$.
The size of the 2D box was varied to ensure the same density as 
for the nanoparticles at liquid/vapor interfaces.
A Langevin thermostat with a damping constant $\Gamma=0.1\tau^{-1}$ 
was used to keep the temperature at $T=1.0\epsilon/k_{\rm B}$. 
The thermostat works as an implicit solvent.
The interaction between the nanoparticles is still given by Eq.~(\ref{ColPotential}) 
since capillary interactions are negligible for nanoparticles 
because of the irrelevance of gravity.\cite{kralchevsky94}
We confirmed this treatment in our simulations by directly 
calculating and visualizing the liquid/vapor interface,
which is flat (except for temporary capillary fluctuations) 
all the way to the contact line on the nanoparticle surface 
and shows no distortion at all with the presence of nanoparticles.
It should be pointed out that for larger particles, capillary interactions
can be taken into account even in 2D simulations by adding
an effective capillary attraction between particles.\cite{bleibel11}

\bigskip\noindent{\bf III. RESULTS AND DISCUSSION}
\bigskip

\noindent {\bf A. Out-of-Plane Fluctuations}
\bigskip

\begin{figure}[htb]
\centering
\includegraphics[width=2.5in]{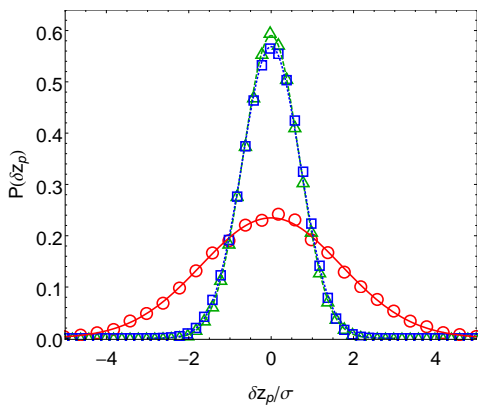}
\caption{(Color online) Probability density distribution $P(\delta z_p)$ of nanoparticle
positions in the $z$-direction, where $\delta z_p$ is the 
deviation of position from the instantaneous mean of all nanoparticles.
Data are for $\phi=0.54$, $\theta_c=93^\circ$,
and various liquids:
monatomic LJ liquid (circles),
10-bead chain polymeric liquid (triangles),
and 100-bead chain polymeric liquid (squares).
Lines are the corresponding Gaussian fits.}
\label{ZFluc}
\end{figure}

At a finite contact angle $\theta_c$, the
nanoparticles straddle the liquid/vapor interface and form 
a layer that is essentially 2D, with small thermal fluctuations
in the $z$-direction normal to  the interface. 
Direct calculation of the magnitude of fluctuations shows that
it is almost independent of $\theta_c$ and the nanoparticle density $\phi$,
but strongly depends on the properties of the supporting liquid film, 
including its density, viscosity, and surface tension.
Because of thermal fluctuations, the layer thickness is broadened 
and the individual nanoparticle height deviates from the mean value
of all nanoparticles.
Such deviations are apparent in snapshots shown in Fig.~\ref{FigSnapshot}.
More quantitative results are shown in Fig.~\ref{ZFluc},
where the probability density distribution $P(\delta z_p)$
is calculated as a function of the deviation $\delta z_p$
from the instantaneous mean height of all nanoparticles.
Note that the mean height itself fluctuates with time,
but such fluctuations are excluded in Fig.~\ref{ZFluc}.
Including them would make the distributions even wider. 
The distributions in Fig.~\ref{ZFluc} all have a Gaussian shape,
with a variance $(1.75\pm 0.08)\sigma$ for 
the monatomic LJ liquid, and $\sim 0.7\sigma$ for
the two polymeric liquids.
Since the nanoparticle diameter is $20\sigma$,
the height variations amongst the nanoparticles 
are less than $10\%$ of their diameter,
which confirms that the nanoparticle layer is close to a 2D system.

\bigskip
\noindent {\bf B. In-Plane Structure}
\bigskip

To characterize the structure in the plane of the 
nanoparticle layer, we computed the 2D radial distribution function $g(r)$
and the structure factor $S({\bf q})$, where
${\bf q} = q_x {\it e_x} + q_y {\it e_y}$ is
a 2D wave-vector and ${\it e_x}$ and ${\it e_y}$
are unit vectors along $x$ and $y$ directions, respectively. 
While $g(r)$ and $S({\bf q})$ are
related through a Fourier transform, for a finite system
it is easier to calculate each one directly. 
The calculation of $g(r)$ involves counting the number of pairs of nanoparticles
separated by distance $r$ and is straightforward.
The structure factor is given by
\begin{equation}
S({\bf q})=N_p^{-1}\sum_{m,n}{\rm exp}(i{\bf q}\cdot{\bf r}_{mn}),
\end{equation}
where the sum is taken over all nanoparticle pairs indexed by $m$ and $n$
and separated by ${\bf r}_{mn}={\bf r}_{m}-{\bf r}_{n}$.

\begin{figure}[htb]
\centering
\includegraphics[width=3.25in]{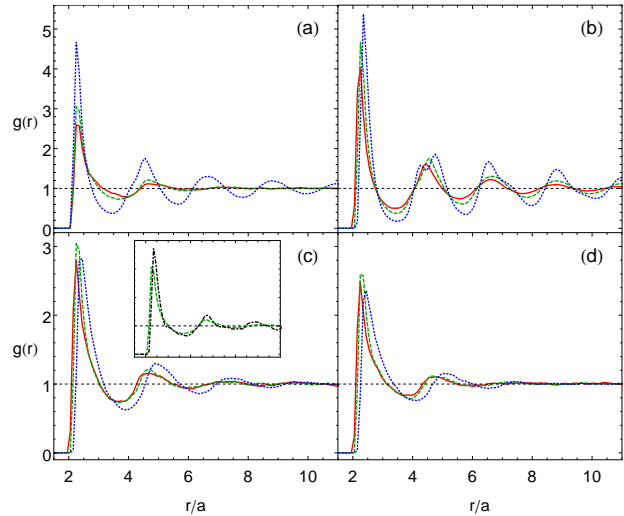}
\caption{(color online) The pair distribution function $g(r)$.
Lines in all main panels are for the monatomic LJ liquid:
(a) $\theta_c=93^\circ$ and 
$\phi=0.45$ (solid), $0.54$ (dashed), $0.73$ (dotted);
(b) $\phi=0.73$ and $\theta_c=137^\circ$ (solid), $93^\circ$ (dashed), $51^\circ$ (dotted);
(c) $\phi=0.54$ and $\theta_c=137^\circ$ (solid), $93^\circ$ (dashed), $51^\circ$ (dotted); 
(d) $\phi=0.45$ and $\theta_c=137^\circ$ (solid), $93^\circ$ (dashed), $51^\circ$ (dotted).
Inset of (b): The dashed line is for the monatomic LJ liquid and
the dash-dotted line is for the polymeric liquid consisting of 100-bead chains; 
for both lines $\phi=0.54$ and $\theta_c=93^\circ$;
}
\label{GofRPlot}
\end{figure}

Results of $g(r)$ are shown in Fig.~\ref{GofRPlot}.
Data in Fig.~\ref{GofRPlot}(a) are for different $\phi$'s at $\theta_c=93^\circ$. 
These results show the expected increase in the height 
of the first peak in $g(r)$ as $\phi$ increases.
The effect of $\theta_c$ on $g(r)$
is illustrated in Fig.~\ref{GofRPlot}(b)-(d).
All results consistently show that when $\theta_c$ is reduced, 
the locations of the peaks of
$g(r)$ move to larger $r$, which indicates that the nanoparticle layer
is slightly denser at larger $\theta_c$.
This trend is consistent with experimental results on the silica nanoparticle layer
at the water-air interface.\cite{zang11}
It can be understood from a simple physical picture.
At large $\theta_c$, the nanoparticles 
ride high at the liquid/vapor interface and the separation 
between nanoparticles is solely controlled by the hard core repulsion.
However, at small $\theta_c$, the nanoparticles are partially
coated by the liquid, which increases their effective size.
So as $\theta_c$ decreases 
the mean separation between two nanoparticles increases.

Figure~\ref{GofRPlot}(b) shows that at $\phi=0.73$ the peaks of $g(r)$ grow
as $\theta_c$ decreases,
indicating that the nanoparticle layer becomes more ordered at smaller $\theta_c$.
The data in Fig.~\ref{GofRPlot}(c) and (d) for $\theta_c=137^\circ$ and 
$93^\circ$ also show this trend. However, the first peak of $g(r)$
becomes lower when $\theta_c$ is further reduced to $51^\circ$,
though other peaks become higher. The apparent reduction in the first peak
of $g(r)$ at $\theta_c=51^\circ$ is due to a finite size effect.
The calculation of $S({\bf q})$ indicates that all peaks grow as $\theta_c$ 
decreases and confirms that the nanoparticle layer exhibits stronger local order
at smaller $\theta_c$.

As $\theta_c$ continues to decrease to $0$, 
we have observed the loss of nanoparticles at the interface as
some of them diffuse into the liquid, 
though $\theta_c$ for an individual nanoparticle may still be finite.
This effect is more significant at higher nanoparticle density.
For example, at $\theta_c=29^\circ$ more than $18\%$ nanoparticles were dispersed in the liquid
at the end of MD runs for $\phi=0.73$;
while for $\phi=0.45$, about $8\%$ have diffused into the liquid.
The loss of nanoparticles at the interface for a finite $\theta_c$ 
is a consequence of their small activation energy,
which is typically at the order of several $k_{\rm B}T$.\cite{bresme07}

Result of $g(r)$ for the polymeric liquid consisting of 100-bead chains 
is included in the inset of Fig.~\ref{GofRPlot}(c),
together with the result for the monatomic LJ liquid at the same $\theta_c$ and $\phi$.
The peaks of $g(r)$ are slightly higher
for the polymeric liquid than those for the monatomic LJ liquid, 
indicating that local order is slightly stronger in the former case.
As shown in Fig.~\ref{ZFluc}, the out-of-layer fluctuation is 
much smaller for the polymeric liquids, i.e., the nanoparticle layer is more 2D-like.
As a consequence, the nanoparticle density is
effectively higher, which leads to higher peaks in $g(r)$ for the polymeric liquids.

\begin{figure}[htb]
\centering
\includegraphics[width=2.75in]{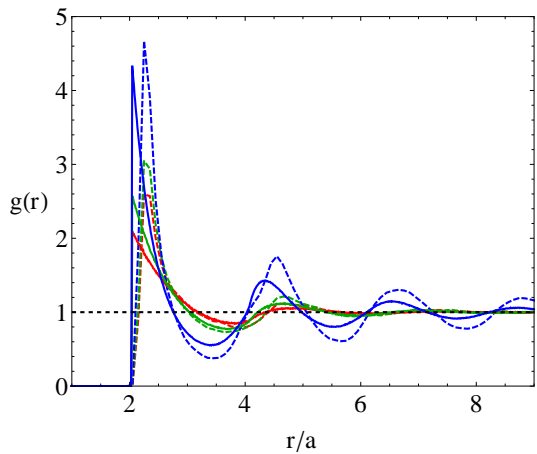}
\caption{(color online) The pair distribution function $g(r)$
for nanoparticles at the liquid/vapor interface for the LJ monatomic liquid 
at $\theta_c=93^\circ$ (dashed lines) compared to results of 2D simulations (solid lines)
for $\phi=0.45$ (bottom), $0.54$ (middle), and $0.73$ (top).}
\label{GofRPlot2D}
\end{figure}

A comparison of $g(r)$ for
nanoparticles at the liquid/vapor interface 
and in 2D with an implicit solvent 
is shown in Fig.~\ref{GofRPlot2D} for three densities. 
The peaks in $g(r)$ for nanoparticles at the interface 
are clearly higher and decay slower than those for nanoparticles in 2D, 
indicating stronger local order in the former case. 
The locations of the peaks move towards larger $r$
for nanoparticles at the interface. 
The shift is between $2$ to $2.5\sigma$,
and reflects the liquid coating that is about $1\sigma$ in thickness
on each nanoparticle.
The coating makes the nanoparticles effectively larger 
than the bare ones in 2D simulations. 
Therefore, the actual nanoparticle density is effectively
higher at the interface, which leads to higher peaks in $g(r)$.
However, if the radius of nanoparticles in 2D was increased to reflect this coating, 
then we would expect higher peaks in $g(r)$ at a given $\phi$ from 2D simulations.
This trend can be derived indirectly from the previous comparison
of $g(r)$ between the monatomic LJ liquid and polymeric liquids.
Higher peaks are found for the latter since the nanoparticle layer there is more 2D-like.

\begin{figure}[htb]
\centering
\includegraphics[width=3.25in]{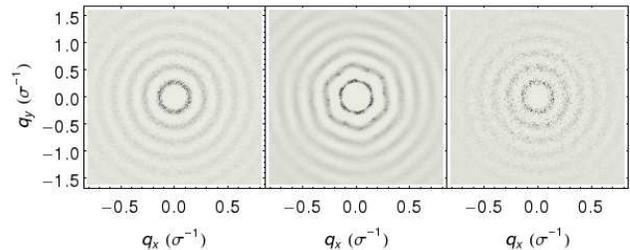}
\caption{(color online) 
Density plots of $S({\bf q})$ in the $q_x$-$q_y$ plane 
for $\theta_c=93^\circ$ and the monatomic LJ liquid
at $\phi=0.54$ (left) and $0.73$ (middle),
and the 100-bead chain polymeric liquid at $\phi=0.54$ (right).}
\label{SofQ2DPlot}
\end{figure}

Additional information on the in-plane structure of the nanoparticle layer can be obtained from $S({\bf q})$. 
A density plot of $S({\bf q})$ in the $q_x$-$q_y$ plane is shown 
in Fig.~\ref{SofQ2DPlot} for $\theta_c=93^\circ$ and the monatomic LJ liquid
at $\phi=0.54$ and $0.73$, and 
the 100-bead chain polymeric liquid at $\phi=0.54$. 
At low density $\phi=0.45$ and $0.54$, the local structure is 
almost isotropic, indicating a fluid-like state of the floating layer.
However, at $\phi=0.73$, which is still lower than the critical density
for the fluid/solid transition of 2D hard sphere systems, 
the hexagonal close-packing feature of the local structure is rather apparent 
as shown in the middle panel of Fig.~\ref{SofQ2DPlot}.
Smaller out-of-plane fluctuations of nanoparticles on the surface of 
the polymeric liquid lead to stronger local order in the nanoparticle layer
at a given $\phi$ and $\theta_c$, indicated by higher peaks in $S({\bf q})$.

\bigskip
\noindent {\bf C. Orientational Order}
\bigskip

A simple measure of orientational order of the floating nanoparticle layer is provided by
the fraction of nanoparticles, $f_6$, which have exactly
six neighbors as in a close-packed hexagonal lattice.
Here a simple criterion is adopted
to identify nearest neighbors as those within
a cut-off radius $r_c$ of a given nanoparticle.
We chose $r_c=35\sigma$, roughly corresponding to the location
of the first valley of $g(r)$ as shown in Fig.~\ref{GofRPlot}.
After nearest neighbors were found, we computed
the Nelson-Halperin (N-H) order parameter, 
which can be expressed as
\begin{equation}
m_6 = \left< \left| \frac{1}{N_p}\sum_{j=1}^{N_p}
\frac{1}{n_j}\sum_{k=1}^{n_j}\exp (i6\theta_{jk})\right| \right>,
\label{NHEq}
\end{equation}
where the first sum is over all $N_p$ nanoparticles,
$n_j$ is the number of nearest neighbors of the $j$-th nanoparticle,
the second sum is over $n_j$ nearest neighbors,
and $\theta_{jk}$ is the angle formed by the bond between a nearest neighbor
pair $(j,k)$ and a fixed axis.
For an ideal hexagonal lattice $m_6=1$, and it
decreases to zero as the local bond disorder increases.

\begin{figure}[htb]
\centering
\includegraphics[width=2.5in]{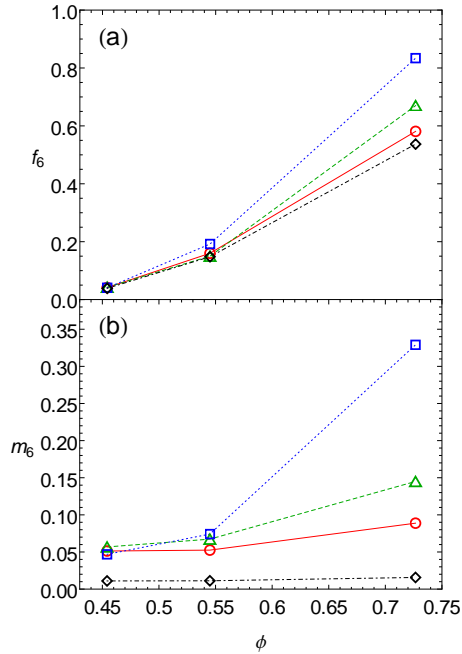}
\caption{(a) The fraction of nanoparticles, $f_6$, having exactly six neighbors and 
(b) the N-H order parameter $m_6$ vs. the nanoparticle density $\phi$
for nanoparticles on the surface of the monatomic LJ liquid at various contact angles:
$\theta_c=137^\circ$ (circles);
$\theta_c=93^\circ$ (triangles);
$\theta_c=51^\circ$ (squares).
Results of 2D simulations are shown with diamonds.
Lines are guides to the eye.}
\label{BondOrderPlot}
\end{figure}

Results of $f_6$ and $m_6$ vs. $\phi$ are shown in Fig.~\ref{BondOrderPlot} 
for nanoparticles floating on the surface of the monatomic LJ liquid 
at various contact angles.
Results of 2D simulations are also included.
As expected, in all cases $f_6$ and $m_6$ increase with $\phi$, indicating 
the development of orientational order.
Figure~\ref{BondOrderPlot} also shows that at a given $\phi$, the layer
is more locally ordered at smaller $\theta_c$ as
indicated by larger values of $m_6$ and $f_6$.
The increase in orientational order as $\theta_c$ is reduced 
is more dramatic at higher $\phi$.
This trend persists in our simulations when $\theta_c$ is reduced
even further (e.g., $\theta_c=29^\circ$).
In this case since some nanoparticles are
eventually absorbed into the liquid at long times, we can only calculate $f_6$ and
$m_6$ in the early stages
of simulations when all nanoparticles are still at the liquid/vapor interface.
In this early time regime systems follow the same trend that at a given $\phi$, 
values of $f_6$ and $m_6$ are larger and the order is stronger at smaller $\theta_c$.
The trend is consistent with that found from $g(r)$ as shown in Fig.~\ref{GofRPlot}.
Data in Fig.~\ref{BondOrderPlot} also show that orientational order
is stronger for nanoparticles at the interface than for those in 2D since
the liquid coating in the former case makes nanoparticles effectively larger
and their density higher. 
The same reason also leads to enhancement of translational order as 
shown in Fig.~\ref{GofRPlot2D}.

The N-H order parameter $m_6$ shown in Fig.~\ref{BondOrderPlot} 
was calculated with nearest neighbors identified with a simple cut-off criterion.
To verify that this criterion leads
to an accurate estimate of orientational order, we also conducted
a Voronoi analysis of the packing geometry of the nanoparticle layer.\cite{steinhardt83} 
Nearest neighbors were identified
as those sharing common sides in the Voronoi construction. 
The N-H order parameter was then computed with Eq.~(\ref{NHEq}).
Results from the Voronoi analysis typically agree 
within a few percent with those from the simpler cut-off approach.
Thus the cut-off criterion is adequate
for finding nearest neighbors needed for the
calculation of the N-H order parameter, 
even when the nanoparticle density is low.

\bigskip
\noindent {\bf D. Diffusion Coefficient}
\bigskip

\begin{figure}[htb]
\centering
\includegraphics[width=2.5in]{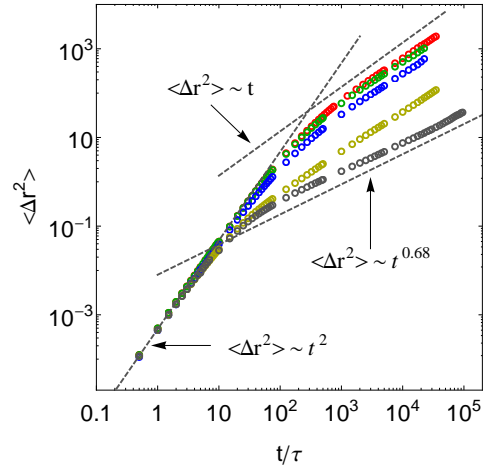}
\caption{(a) Mean square displacement $\langle \Delta r^2 \rangle$ vs. time $t$.
The top three data sets are for the monatomic LJ liquid at 
$\phi=0.45$ and $\theta_c=137^\circ$;
$\phi=0.54$ and $\theta_c=137^\circ$;
$\phi=0.54$ and $\theta_c=93^\circ$.
The second to bottom (bottommost) data set is for the polymeric liquid consisting of
10-bead (100-bead) chains at $\phi=0.54$ and $\theta_c=93^\circ$.}
\label{MSQDPlot}
\end{figure}

\begin{figure}[htb]
\centering
\includegraphics[width=2.5in]{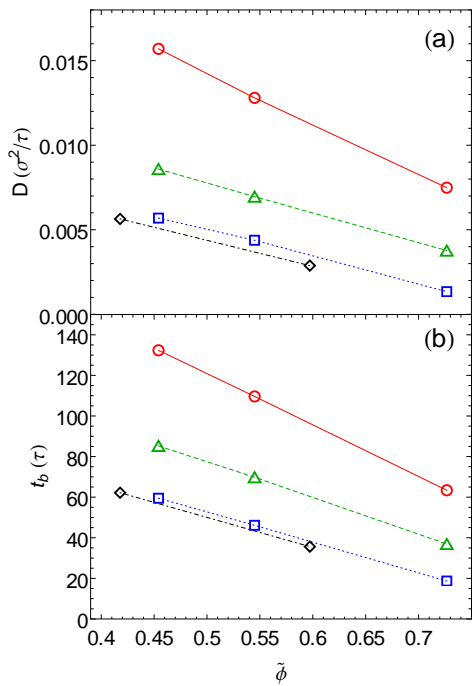}
\caption{(a) Diffusion coefficient $D$ and (b) ballistic time $t_b$ 
vs. the true nanoparticle density $\tilde{\phi}$ at various contact angles:
$\theta_c=137^\circ$ (circles),
$\theta_c=93^\circ$ (triangles),
$\theta_c=51^\circ$ (squares),
$\theta_c=29^\circ$ (diamonds).
$\tilde{\phi}$ is the same as $\phi$ except for $\theta_c=29^\circ$,
where $\tilde{\phi}<\phi$.
Lines are guides to the eye.}
\label{DiffConstPlot}
\end{figure}

The in-plane motion of nanoparticles at the liquid/vapor interface
is characterized by the mean square displacement 
$\langle \Delta r^2 \rangle \equiv \langle \Delta x^2 +\Delta y^2 \rangle$, which
is shown in Fig.~\ref{MSQDPlot} for various cases.
At very short times, the motion of nanoparticles is
nearly ballistic with $\langle \Delta r^2 \rangle = v^2t^2$, where
$v$ is a velocity and $t$ is time. 
At larger times, 
the motion becomes diffusive and $\langle \Delta r^2 \rangle = 4Dt$, 
where $D$ is a diffusion coefficient and 4 is the prefactor for 2D diffusion.
Most of our results fit to this classical picture, 
as shown in Fig.~\ref{MSQDPlot}.
In all cases, $v \simeq 0.02\sigma/\tau$ and is nearly independent
of $\theta_c$ and $\phi$.
The intersection of the ballistic and diffusive regime defines a ballistic time scale $t_b=4D/v^2$. 
Results for $D$ and $t_b$ are shown in Fig.~\ref{DiffConstPlot}
as a function of the true nanoparticle density $\tilde{\phi}$ 
for 4 values of $\theta_c$. At small $\theta_c$ (e.g., $\theta_c=29^\circ$), because of
the loss of nanoparticle from the interface, $\tilde{\phi}$ is lower
than the nominal density $\phi$ calculated from $N_p$.
In other cases $\tilde{\phi} = \phi$.
Figure~\ref{DiffConstPlot} shows that in all cases, $D$ and $t_b$ decrease approximately linearly
with the nanoparticle density and the slope is steeper for larger $\theta_c$. 
At the same $\tilde{\phi}$, both $D$ and $t_b$ decrease as $\theta_c$ decreases. 
This is due to the fact that 
the attraction between the nanoparticles and liquid is enhanced to make $\theta_c$ smaller.
As a consequence, nanoparticles become more immersed into the liquid 
and need to plow through more liquid in order to move,
which makes their diffusion more difficult. 
At $\tilde{\phi}=0.45$, $D$ and $t_b$ decrease
by a factor of $3$ when $\theta_c$ is reduced from $137^\circ$
to $29^\circ$. At $\tilde{\phi}=0.73$,
the reduction is almost $6$-fold for the same change in $\theta_c$.

For the 10-bead chain polymeric liquid, 
$\langle \Delta r^2 \rangle$ grows linearly with $t^2$ at early times 
and shows diffusive behavior at large times. 
However the diffusion coefficient is reduced significantly compared to the LJ monomer case.
For example, at $\tilde{\phi}=0.54$ and $\theta_c=93^\circ$, the value
of $D$ is  $\sim 1\times 10^{-3}\sigma^2/\tau$ for the 10-bead polymeric liquid 
compared to $\sim 7\times 10^{-3}\sigma^2/\tau$ for the monatomic LJ liquid.
Note that Fig.\ref{FigConAng} shows that the interaction strength 
between the nanoparticles and liquid, 
dictated by $A_{\rm ns}$, has to be adjusted to ensure the same $\theta_c$
for the polymeric and monatomic LJ liquid. 
For example, $\theta_c=93^\circ$ for the monatomic LJ liquid at $A_{\rm ns}=60\epsilon/\sigma^2$,
while we need to increase $A_{\rm ns}$ to $79\epsilon/\sigma^2$ to get the same $\theta_c$
for the 10-bead polymeric liquid.
If $A_{\rm ns}$ is held fixed at $A_{\rm ns}=60\epsilon/\sigma^2$ for the 10-bead polymeric liquid,
then $\theta_c$ increases to $113^\circ$. 
In this case $D$ is $\sim 1.6\times 10^{-3}\sigma^2/\tau$.
These data are summarized in Table~\ref{diffdata}.
If we compare the values of $D$ between the monatomic LJ and
10-bead polymeric liquid at $A_{\rm ns}=60\epsilon/\sigma^2$, 
i.e., at the same nanoparticle-bead interaction,
we find that $D$ is reduced by a factor about $4$.
Note that the viscosity $\eta$ of the 10-bead polymeric liquid is 
approximately $12$ times larger than that of the monatomic LJ liquid.
If we assume a scaling relation between $D$ and $\eta$ as $D \sim \eta ^{-\alpha}$,
then $\alpha \simeq 0.6$ is much less than 1, which
is in contrast with the situation of 
free bulk Brownian diffusion where $\alpha = 1$ is expected.
However, our results are consistent with a recent experimental measurement of
the diffusion coefficient of single nanoparticles at water-oil interfaces,
where $\alpha = 0.44$ was found.\cite{wang11}

\begin{table}
\begin{tabular}{|c|c|c|c|c|}
\hline
liquid & $A_{\rm ns}$ ($\epsilon/\sigma^2$) & $\theta_c$ ($^\circ$) 
& $\eta$ ($m/\sigma\tau$) & $D$ ($\sigma^2/\tau$) \\
\hline
monatomic & 60 & 93 & 1.01 & $7.0\times 10^{-3}$ \\
\hline
10-bead & 60 & 113 & 12 & $1.6\times 10^{-3}$ \\
\hline
10-bead & 79 & 93 & 12 & $1.0\times 10^{-3}$ \\
\hline
\end{tabular}
\caption{Comparison of values of contact angle ($\theta_c$), 
viscosity ($\eta$), and diffusion coefficient ($D$) at $\phi=0.54$
for the monatomic LJ and 10-bead polymeric liquid.}
\label{diffdata}
\end{table}

For the 100-bead chains, $\langle \Delta r^2 \rangle \sim t^2$ at early times, 
and then crosses over to a sub-diffusive regime as shown in Fig.~\ref{MSQDPlot}. 
This deviation from the expected long-time linear dependence on time 
continued out to the longest times we are presently able to simulate. 
In this sub-diffusive regime, the mean square displacement for the 100-bead chains
can be better fit to $\langle \Delta r^2 \rangle \sim t^{0.68}$, as shown in Fig.~\ref{MSQDPlot}.
The reason underlying this sub-diffusive regime is unclear.

\bigskip
\noindent{\bf V. CONCLUSIONS}
\bigskip

In this paper we studied nanoparticles floating at liquid/vapor interfaces with MD simulations. 
Both the low-viscous monatomic LJ liquid and the high-viscous polymeric liquids composed
of flexible linear chains were studied.
We showed that as the attraction between the nanoparticles and liquid
is increased, the contact angle is reduced 
and the nanoparticles are more wetted by the liquid.
At the same time the short range order of the nanoparticle layer is slightly enhanced and
the interfacial diffusion coefficient of the nanoparticles is greatly reduced.
Our results further showed that both the translational and orientational order
of the nanoparticle layer
grow quickly and the nanoparticle diffusion slows down dramatically 
as the nanoparticle density is increased.
Comparisons with results of 2D simulations revealed that the main effect of liquid 
on the nanoparticles is to provide a coating 
which makes their effective size larger than that of the bare ones.
Otherwise, the nanoparticle layer at the liquid/vapor interface is close to a 2D system,
though the out-of-plane fluctuations can be as large as $10\%$ of their diameter.
The simulations with more viscous polymeric liquids showed that the
the out-of-plane fluctuations of the nanoparticles are strongly suppressed 
even for relatively short chains (e.g., 10-bead chains).
The local, short range order is slightly enhanced at a given contact angle 
and nanoparticle density as the chain length increases.
At the same time, the nanoparticle diffusion becomes slower for more viscous liquids and 
even shows sub-diffusive behavior at large times for highly viscous liquids.
The diffusion coefficient scales inversely with the viscosity however with an exponent less than 1.

\bigskip
\noindent{\bf ACKNOWLEDGMENTS}
\bigskip

This research used resources of the National Energy Research Scientific 
Computing Center (NERSC), which is supported by the Office of Science 
of the United States Department of Energy 
under Contract No. DE-AC02-05CH11231, and the Oak Ridge Leadership Computing Facility 
located in the National Center for Computational Sciences at Oak Ridge National Laboratory, 
which is supported by the Office of Science of the United States Department of Energy 
under Contract No. DE-AC05-00OR22725. 
These resources were obtained through the Advanced Scientific Computing Research (ASCR) 
Leadership Computing Challenge (ALCC).
This work was supported by the Laboratory Directed Research and 
Development program at Sandia National Laboratories.
Sandia National Laboratories is a multi-program laboratory 
managed and operated by Sandia Corporation, 
a wholly owned subsidiary of Lockheed Martin Corporation, 
for the U.S. Department of Energy's National Nuclear Security Administration 
under contract DE-AC04-94AL85000.


\begin{thebibliography}{47}
\bibitem[{not({\natexlab{a}})}]{note1}
\bibinfo{note}{As results for the shear viscosity for $N=10$ and 100 were not
  known for $r_c=3.0\sigma$, we equilibrated systems of 25000 chains of length
  $N=10$ and 2500 chains of length $N=100$ at zero pressure and then sheared at
  a given shear rate. Non-equilibrium trajectories were generated by
  integrating the SLLOD equations of motion.\cite{tuckerman97} The shear
  viscosity $\eta$ was calculated for several shear rates and extrapolated to
  zero shear rate. We found $\eta=12\pm 1$ and $180\pm 10~m/\sigma\tau$ for
  $N=10$ and $100$, respectively.}
\bibitem[{not({\natexlab{b}})}]{note3}
\bibinfo{note}{Calculating $\theta_c$ as a function of $A_{\rm ns}$ provides a
  convenient way to determine the phase diagram of a nanoparticle/liquid
  mixture. Studies along this line will be presented in a future publication.}
\end{thebibliography}

\end{document}